\documentclass{IEEEtran}

\usepackage{multicol}
\usepackage{lipsum}
\usepackage{xcolor}
\usepackage{multirow}
\usepackage{enumitem}
\usepackage{geometry}

\usepackage{amsmath,amsfonts}
\usepackage{algorithmicx}
\usepackage{algorithm}
\usepackage{algpseudocode}
\usepackage{array}
\usepackage{caption}
\usepackage{textcomp}
\usepackage{stfloats}
\usepackage{url}
\usepackage{verbatim}
\usepackage{cite}
\usepackage{color}
\usepackage{booktabs}
\usepackage{makecell}
\usepackage{graphicx}
\usepackage{enumitem}
\usepackage{multirow}
\usepackage{footnote}
\usepackage{enumerate}
\usepackage{pifont}

\makeatletter
\def\footnoterule{\kern-3\p@
	\hrule \@width 0.49\textwidth \kern 2.6\p@}
\makeatother

\usepackage{caption}
\newcommand{\subparagraph}{}
\usepackage{titlesec}
\titlespacing{\section}
{0pt}{5pt}{5pt}
\titlespacing{\subsection}
{0pt}{5pt}{5pt}

\begin{document}

\begin{flushleft}
	\fontsize{14}{15} \fontfamily{phv}\selectfont
	 \textbf{BandMap: Application Mapping with Bandwidth Allocation for Coarse-Grained Reconfigurable Array
	}
\end{flushleft}

\begin{flushleft}	
	{\fontsize{10}{10} \fontfamily{phv}\selectfont 
		Xiaobing Ni\textsuperscript{1}, JiaHeng Ruan\textsuperscript{1}, Mengke Ge\textsuperscript{1,2}, Wendi Sun\textsuperscript{1}, Song Chen\textsuperscript{1,2},  Yi Kang\textsuperscript{1,2} \\
		\textsuperscript{1}School of Microelectronics, University of Science and Technology of China\\
		\textsuperscript{2}Institute of Artificial Intelligence, Hefei Comprehensive National Science Center
	}

\end{flushleft}
	\noindent
	\begin{abstract}
		 This paper proposes an application mapping algorithm, BandMap, for coarse-grained reconfigurable array (CGRA), which allocates the bandwidth in PE array according to the  transferring demands of  data, especially  the data with high spatial reuse, 
		 to reduce the routing PEs. 
		 To realize the application mapping with bandwidth allocation,
		 BandMap  maps the data flow graphs (DFGs), abstracted from applications' kernel loops, onto CGRA by solving the maximum independent set (MIS) on a mixture of tuple and quadruple resource occupation conflict graph. 
		 Compared to a state-of-the-art BusMap work, Bandmap can achieve reduced routing PEs with the same or even smaller initiation interval (II).
		
	\end{abstract}
	
	\noindent

\section{Introduction}
Compute-intensive applications share the common properties of high computing intensity, streaming memory access and data reuses\cite{nowatzki2017stream}, which  grows explosively in broad fields, e.g. AI, DSP, etc. The CGRA in Fig.1, featuring abundant parallel computing resources, high energy efficiency, flexibility and the ability to provide streaming dataflow\cite{nowatzki2017stream} for the PE array (PEA), is a promising spatial architecture for these applications\cite{BusMap}.

The data  reuses in the applications occur temporally and spatially  during the computing process. For spatial reuse,  the data, typically the input data, has to be transferred to different PEs simultaneously. However, due to the constraint of the CGRA's  topology and interconnection structure, routing PEs may be needed to cache the data, and broadcast the  reused data to the computing PEs, as Fig.2(a)(b)(d) shows. The occupations of routing PEs lead to power costs, and may even increase the II, thereby reducing the CGRA throughput. 

To reduce the routing PEs in spatial data reuse,  as shown in Fig.2(c)(e), we propose to allocate the bandwidth resources of the PEA considering the data transferring demands. A  mapping algorithm with bandwidth allocation, BandMap, is proposed to resolve the mapping of DFG onto CGRA by solving MIS on a mixture of tuple and quadruple resource occupation conflict graph. Compared to BusMap\cite{BusMap}, we can achieve fewer routing PEs with the same or even smaller II.

\section{CGRA Architecture}
The CGRA in Fig.1 is composed of a 2-D PEA with dimensions $M$$\times$$N$, a host controller, a context memory, a crossbar, and data memories. The PEA exploits  input and output buses denoted as IBUS and OBUS to transfer the input and output data between PEA and memories via input and output ports, denoted as IPORT and OPORT.
The buses and ports constitute the  bandwidth resources of the PEA. 
The PEs attached to a common bus can receive the same data simultaneously.

To implement bandwidth allocation, we  model the input and output data as virtual input and output operations(denoted as \emph{\textbf{VIOs}} and \emph{\textbf{VOOs}} respectively), and bind the virtual operations with ports. Especially for the VIO,  multiple ports binding can be realized by  the crossbar with  multicast configuration.

\begin{figure}[t]
	\centering
	\centerline{\includegraphics[width=0.8\linewidth]{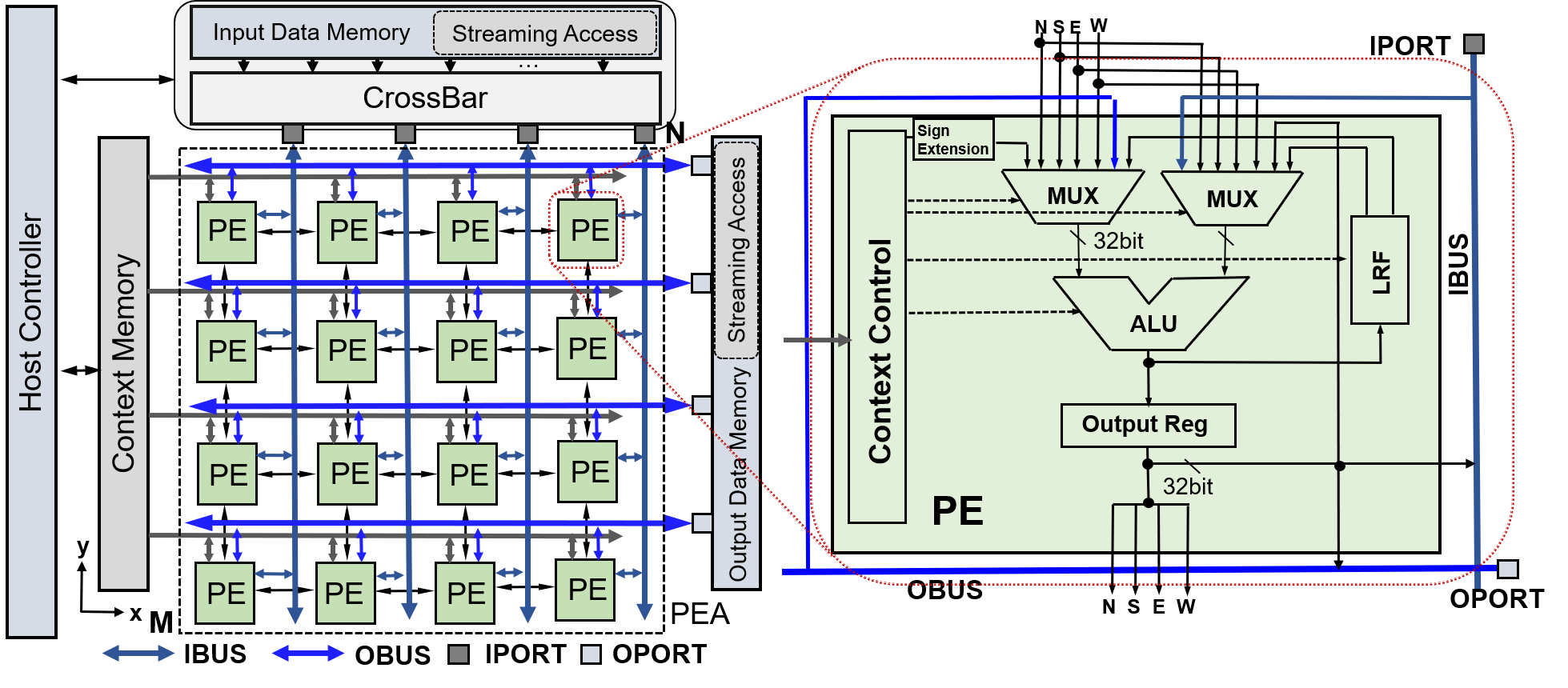}}
	\captionsetup{font=footnotesize}\caption{CGRA architecture}
	\vspace{-0.2cm}
\end{figure}
\begin{figure}[t]
	\centering
	\centerline{\includegraphics[width=0.9\linewidth]{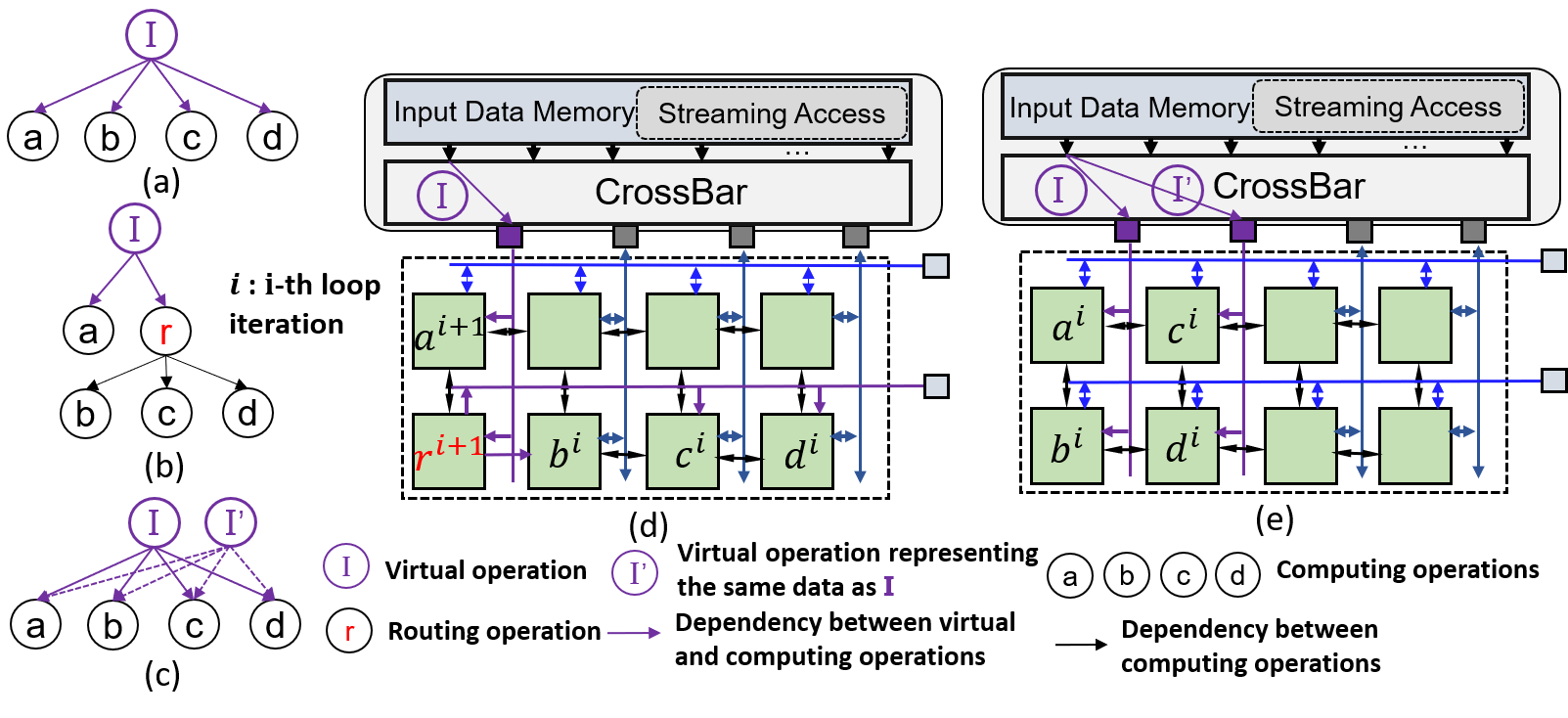}}
	\captionsetup{font=footnotesize}\caption{(a) Virtual operation $I$ is reused by four computing operations $a$, $b$, $c$ and $d$; (b) Adding routing operation; (c) Adding virtual operation; (d) Broadcasting from routing PE; (e) Broadcasting by multiple ports}
	\vspace{-0.0cm}
\end{figure}

\begin{figure}[t]
	\centering
	\centerline{\includegraphics[width=0.9\linewidth]{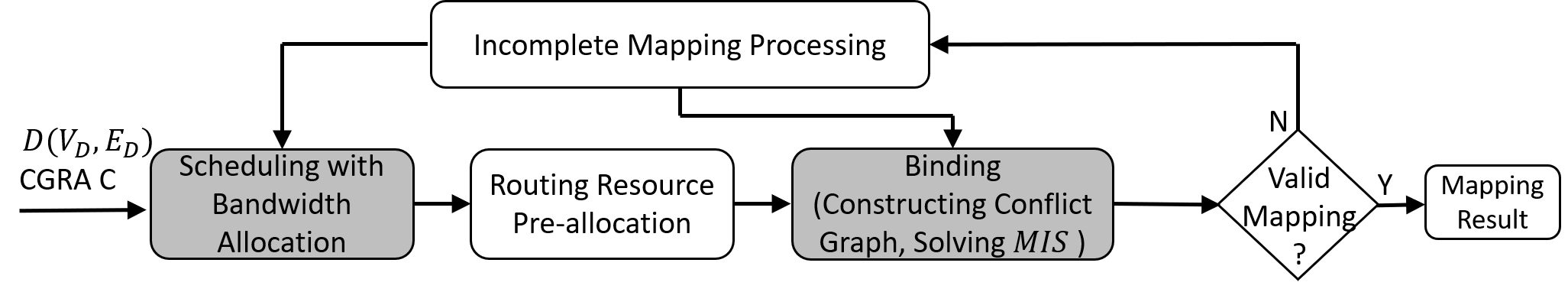}}
	\captionsetup{font=footnotesize}\caption{The overview of BandMap.}
	\vspace{-0.3cm}
\end{figure}

\begin{figure}[t]
	\centering
	\centerline{\includegraphics[width=0.9\linewidth]{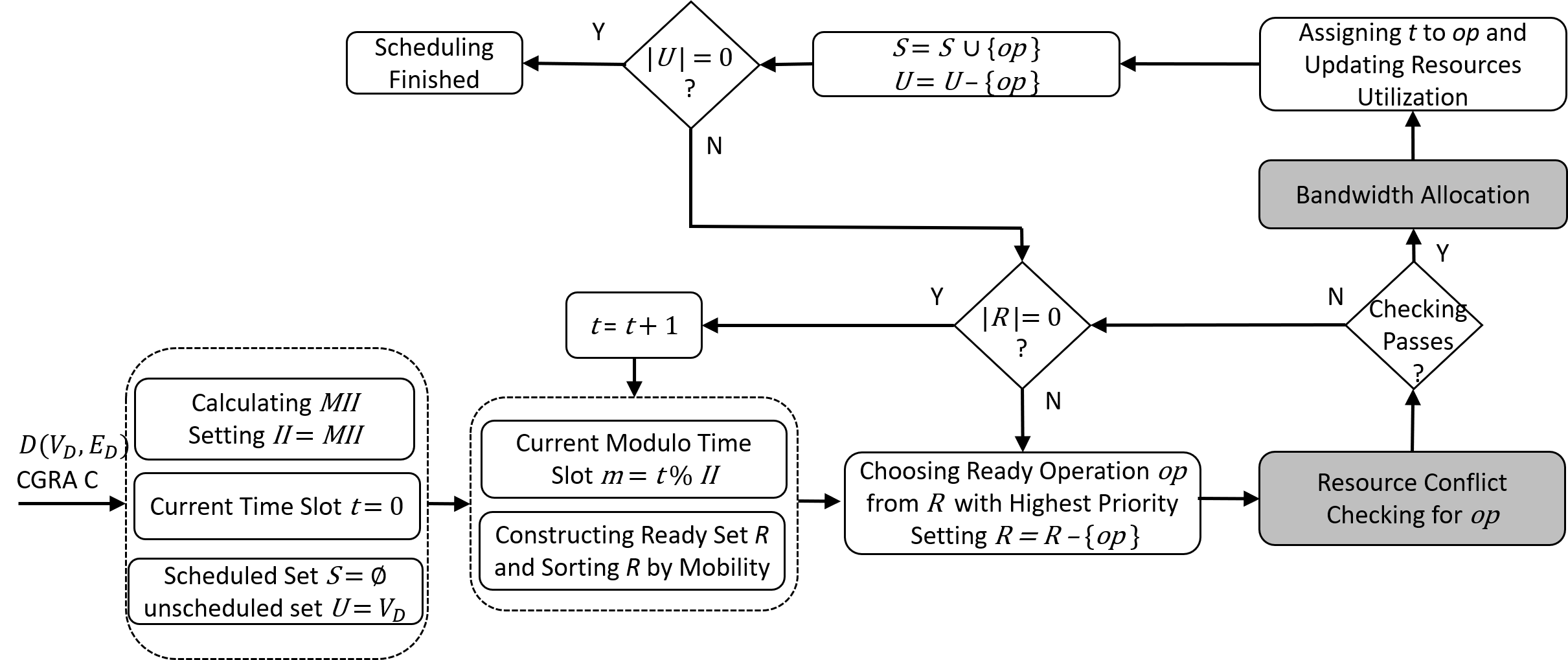}}
	\captionsetup{font=footnotesize}\caption{The overview of scheduling procedure.}
	\vspace{-0.5cm}
\end{figure}

\begin{table}[!htp]
	\vspace{-0.2cm}
	\centering
	\captionsetup{font=footnotesize}\caption{List of Notations}
	\resizebox{1.0\columnwidth}{!}{
		\begin{tabular}{p{1.3cm}p{8.00cm}}
			\toprule
			Notation & Description\\
			\toprule
			$II$     & {Time slots between two successive iterations of  kernel loop.} \\
			$MII$     & {Minimum II, equaling to $max$(resource-constrained MII, recurrence-constrained MII) \cite{Modulo}.} \\
			$D(V_D,E_D)$ & {DFG, wherein $V_D =V_r \cup V_s$, $E_D =E_r \cup E_s$.} \\ 
			$V_r,E_r$ & {$V_r$: computing operations set; $E_r$: directed edge set composed of  the dependencies between computing operations.}\\
			$V_s,E_s$ & {$V_s = V_i\cup V_o$, composed of virtual operations  where $V_i$ and $V_o$ are virtual input and output operation sets; $E_s$, directed edge set composed of  dependencies between virtual and computing operations.}\\
			$RD(op)$ &  Spatial reused degree of of operation $op$ $\in$ $V_s$.\\
			$T_{II}(V_T, E_T)$ & {Time-Extended CGRA (TEC),  replicating  CGRA from 0 to II -1; $V_T$: vertex set composed of replicated nodes. $E_T$: directed edge set where each edge indicates a routing path.  }\\
			$CG(V_{C}, E_{C})$& {$CG$: conflict graph constructed among scheduled DFG and TEC. $V_{C}$: vertex set comprised of  tuples  $(port^{t}_n,$$ op^t_s)$ and quadruples  $(pe^{t}_{i,j},$$ op^t_r,$$ bus^t_{i,x}, $$bus^t_{j,y})$. $port^{t}_n$  denotes  the $n$-th  port instance at time $t$ and $op_s$ $\in$ $V_s$  is transferred via $port^{t}_n$ at time $t$. The definition of $(pe^{t}_{i,j},$$ op^t_r,$$ bus^t_{i,x}, $$bus^t_{j,y})$ can be referred to \cite{BusMap}. $E_{C}$: edge set wherein each edge indicates a conflict relation between two vertices which cannot coexist in any independent set of $CG$. 
			}\\
			$MIS$ & {The $MIS$ in $CG$ means the most operations mapped onto TEC without resource occupation conflicts.}\\
			\bottomrule
		\end{tabular}
	}\vspace{-0.4cm}
\end{table}

\section{BandMap algorithm}
We propose a DFG mapping algorithm with bandwidth allocation, BandMap, to solve the mapping with reduced routing PEs  by solving MIS on a mixture of tuple and quadruple conflict graph.  BandMap contains   four phases: (1) scheduling; (2) routing resource pre-allocation; (3)  binding; (4) incomplete mapping processing, as shown in Fig. 3. This paper focuses on phases (1) (3), and  applies  mechanisms described in \cite{BusMap} for (2) (4). We list  the notations in TABLE \uppercase\expandafter{\romannumeral 1} that will appear in the following texts.
 
\subsection{Scheduling with Bandwidth Allocation}
The scheduling phase is in charge of two aspects: assigning scheduling time and allocating bandwidth quantitatively. Since CGRA throughput is the top concern in most applications, to minimize the II, we first schedule the DFG under II = MII, as Fig. 4 shows. The key procedure is  resource conflict checking, i.e., checking whether the PEs (input/output ports)  are available or not for the ready computing operation (VIO/VOO)  at modulo time slot $m$ = $t$ \% II. If checking passes, the operation is assigned with scheduling time $t$ and occupies  one of the PEs (input/output ports) located at TEC's $m$-th time layer. Otherwise, the next ready operation is chosen from the ready set and checked repeatedly.

Since the input data should be immediately transferred to computing PEs, for scheduling the VIO, we also need to check the available PEs at the current modulo time $m$ and determine the allocation of input bandwidth (ports) for VIO based on the reused degree of VIO, $RD$(VIO), and the number of PEs attached to the common IBUS, $M$. We adopt a straightforward allocation policy: At current modulo time $m$, if $RD$(VIO) is greater than $M$, we allocate the VIO with the input ports whose quantity $Q$  equals to $min$($\lceil \frac{RD(VIO)}{M} \rceil$, the number of available input ports). 
If $Q$ $<$ $\lceil \frac{RD(VIO)}{M} \rceil$, or the number of available PEs is smaller than  $RD$(VIO),  routing PEs are adopted. To better model the binding of VIO with multiple ports, we create new $Q$ - 1 VIOs representing the same data as the original VIO, as shown in Fig.2(c)(e), and each VIO occupies one port. 

The output data represented by VOO has no spatial data reuse and occupies one OPORT. The output bandwidth allocation for VOOs is completed in the resource  conflict checking.

\subsection{Binding with Conflict Graph}
The binding phase is responsible for binding the scheduled DFGs to TEC. To cover the bandwidth allocation, we exploit the  tuples  to represent the port occupations and formulate the binding as solving the MIS on a mixture of tuples and quadruples \cite{BusMap} conflict graph $CG(V_C, E_C)$.

The tuples in $V_C$ are composed of IPORT and OPORT occupations, represented as \{$(port_n^t, op_i^t)$ $|$ $\forall$ $op_i$ $\in$ $V_i$; $n$ = 1, ... , $N$\} $\cup$ \{$(port_m^t, op_o^t)$ $|$ $\forall$ $op_o$ $\in$ $V_o$; $m$ = 1, ... , $M$\}.

The edges in $CG$ indicate the resource occupation conflicts between two vertices. Since there are two types of vertices (tuples and quadruples) in $V_C$,  we have to consider the  creation of edges between (1) any two tuples; (2) any tuple and any quadruple; (3) any two quadruples:

(1) At any time, if any port is occupied by any two virtual operations, or any virtual operation occupies any two ports, then conflict occurs and an edge is created between these two tuples.

(2) At any time, if any port is occupied, the bus connected with this port is used for bus routing \cite{BusMap}, or if any input (output) port is occupied, the PE consuming (producing) this data is not  bus connected with this port, then conflict occurs and an edge is created between this tuple and quadruple;

(3) If any two quadruples have resource occupation conflict, an edge is created between these two quadruples \cite{BusMap}.

We apply heuristic SBTS \cite{SBTS} to solve MIS to get the most operations mapped onto TEC. If $|MIS|$ = $|V_D|$, a valid mapping is found, otherwise incomplete mapping happens, then the mapping procedure enters incomplete mapping processing.

\section{Evaluation}
\subsection{Experiment Setup}
We select seven kernel loops, denoted as $C_nK_m$, from convolutional neural networks. In every iteration, $C_nK_m$ consumes $n$ input channels data and produces $m$ output channels data  where each  of  $n$ channel data is  spatially reused by $m$ kernels. 

BandMap is evaluated against  BusMap on the CGRA configured with 4$\times$4 PEA;  a local register file (LRF) with 8 capacity in each PE. Meanwhile, we also compare these two techniques when a  global register file (GRF) with  8 capacity is introduced into CGRA. The GRF can be accessed by all the PEs in parallel.

\subsection{Results and Analysis}
\hspace{-0.35cm}\textbf{Throughput comparison.} The ratio of the MII to the  realized II  indicates the acceleration effect. The higher the ratio, the better throughput is achieved. Without GRF, as shown in Fig.5, BandMap can realize the corresponding MIIs for most  applications, except for $C_3K_6$ and $C_5K_5$.  But for these two applications, we can also achieve the same or better throughput because of the reduced routing PEs caused by our method. With GRF, we can achieve the MIIs for all applications.

\begin{figure}[t]
	\centering
	\centerline{\includegraphics[width=0.9\linewidth]{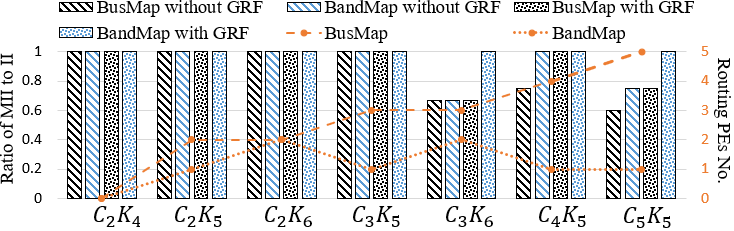}}
	\captionsetup{font=footnotesize}\caption{The mapping results on different $C_nK_m$.}
	\vspace{-0.5cm}
\end{figure}

\hspace{-0.35cm}\textbf{The number of routing PEs comparison.} For application $C_2K_4$, both BusMap and our method need no routing PEs to finish the mapping,  while for the applications with high spatial data reuse ($m > 4$), with bandwidth allocation, our method can reduce the routing PEs by 57.9\% on average and 80\% at most.

\section{Conclusion}
This paper presents BandMap, an application mapping method  for CGRA, which allocates the bandwidth resources in the PEA according to the data transferring demands. The experiment results show BandMap can achieve reduced routing PEs while having the same or even better throughput compared to the state-of-the-art mapping technique.

\ifCLASSOPTIONcaptionsoff
  \newpage
\fi


\end{document}